\documentclass[twocolumn,aps,prb,superscriptaddress]{revtex4}
\usepackage{graphics,bm}
\usepackage{amssymb}
\usepackage{epsfig}
\usepackage{epsf}

\def\be{\begin{equation}} \def\ee{\end{equation}}
\def\beq{\begin{eqnarray}} \def\eeq{\end{eqnarray}}

\def\nn{\nonumber}

\begin{document}

\title{Revisit of Antiferromagnetism in Hubbard Model by A Cluster Slave-Spin Method}
\author{Wei-Cheng Lee}
\email{wlee@binghamton.edu}
\affiliation{Department of Physics, Applied Physics, and Astronomy, Binghamton University - State University of New York, Binghamton, USA}

\author{Ting-Kuo Lee}
\affiliation{Institute of Physics, Academia Sinica, Nankang Taipei 11529, Taiwan}

\date{\today}

\begin{abstract}
The cluster slave-spin method is introduced to systematically investigate the solutions of the Hubbard model including the symmetry-broken phases. 
In this method, the electron operator is factorized into a fermioninc spinon describing the physical spin and a slave-spin describing the charge fluctuations.
Following the $U(1)$ formalism derived by Yu and Si [Phys. Rev. B {\bf 86}, 085104 (2012)], it is shown that the self-consistent equations to explore various symmetry-broken density wave states
can be constructed in general with a cluster of multiple slave-spin sites. 
We employ this method to study the antiferromagnetic (AFM) state in the single band Hubbard model with the two and four-site clusters of slave spins.
While the Hubbard gap, the charge gap due to the doubly-occupied states, scales with the Hubbard interaction $U$ as expected, the AFM gap $\Delta$, 
the gap in the spinon dispersion in the AFM state, exhibits a crossover from the weak to strong-coupling behaviors as $U$ increases. 
Our cluster slave-spin method reproduces not only the traditional mean-field behavior of $\Delta\sim U$ in the weak-coupling limit, but also the behavior of $\Delta \sim t^2/U$ predicted by 
the superexchange mechanism in the strong coupling limit.
In addition, the holon-doublon correlator as functions of $U$ and doping $x$ is also computed, which exhibits a strong tendency toward the holon-doublon binding in the strong coupling regime.
We further show that the quasiparticle weight obtained by the cluster slave-spin method is in a good agreement with the generalized Gutzwiller approximation in both AFM and paramagnetic states, and 
the results can be improved beyond the generalized Gutzwiller approximation as the cluster is enlarged from a single site to 4 sites.
Our results demonstrate that the cluster slave-spin method can be a powerful tool to systematically investigate the strongly correlated system.
\end{abstract}

\maketitle

\section{Introduction}
One critical issue in the Mott physics is how to describe the evolution of physical properties of correlated systems from the weak to strong-coupling limits.\cite{leermp2006}
For example, the single band Hubbard model on the two-dimensional square lattice exhibits an antiferromagnetic (AFM) order with the wavevector $\vec{Q}=(\pi,\pi)$ near the half-filling, but 
the mechanism for the AFM state is fundamentally different in the small and large-$U$ limits, where $U$ is the onsite Coulomb interaction.
In the small-$U$ limit, the normal state can be described by the Fermi liquid theory with well-defined quasiparticles.
Consequently, the AFM mechanism is due to the Fermi surface nesting, and the traditional mean-field picture can be applied to understand the AFM state in the small-$U$ limit.
In the large-$U$ limit, however, the Fermi liquid picture is invalidated due to the strong local charge fluctuations, and it can be shown by a second order perturbation theory 
that the formation of singlet states between nearest-neighbor electrons is energetically favorable.\cite{leermp2006}
As a result, the AFM state can still occur, but the interaction fostering the AFM state is an effective AFM
coupling $J=4t^2/U$ between electron spins on the nearest-neighbor sites, which is known as the superexchange mechanism.\cite{anderson1950}
 
It has been a theoretical challenge to describe the physics in both limits within a single framework. 
Numerical approaches including the exact diagonalization,\cite{dagotto1992,tohyama1994,gooding1994,leung1997,leung2002} 
quantum Monte Carlo method,\cite{bulut1994,varney2009,scalapino2014} and variational Monte Carlo method\cite{leetk1997,leetk2003,leewc2003,wuhk2017} can treat the Hubbard interaction 
non-perturbatively, but they are usually limited to finite-size systems and the extrapolation to the thermodynamic limit is not trivial.
Another line of thinking is to generalize the concept of spin-charge separation in the 1D Hubbard model to higher dimensions.
Theoretical efforts of this sort have been made based on the slave-particle approach.\cite{barnes1976,coleman1984,kotliar1986,florens2004,leermp2006}
The spirit of the slave-particle approach is to introduce auxiliary degrees of freedom to decouple the electron creation and annihilation operators into charge and spin sectors,
and consequently a set of mean-field equations can be derived based on these auxilary degrees of freedom.
Because the dimension of the Hilbert space corresponding to these auxiliary degrees of freedom is larger than the dimension of the original Hilbert space of electrons, the constraint 
to project out unphysical Hilbert spaces has to be considered as one solves the mean-field equations. 
While these slave-particle approaches offer new insights into the quantum correlations due to the Hubbard interaction analytically, 
the main issue is that different choices of the auxiliary degrees of freedom 
as well as the mean-field decoupling schemes can yield quite different results. 
One way to justfy the quality of the solution is to compare them with well-controlled theories in the weak (e.g. Fermi liquid) and strong-coupling (e.g. Gutzwiller approximation) 
limits \cite{kotliar1986,florens2004} or with the dynamical mean-field theory (DMFT) which is exact in the limit of infinite spatial dimensions.\cite{dmft1,dmft2}

Recently, a new slave particle technique called 'slave-spin method' has been widely used to study the orbital-selective Mott transition in multiorbital systems.
\cite{demedici2005,demedici2009,hassan2010,demedici2011,yu2011,yu2012,yu2013,demedici2014,giovannetti2015,mukherjee2016}
Compared to other popular slave-particle approaches,
the advantage of the slave-spin method is that the enlarged Hilbert space due to the slave-spin has a finite dimension, which allows us to treat quantum fluctuations more accurately even at the level 
of the saddle point approximation.
Surprisingly, it has been shown that the slave-spin formalism can obtain the Mott insulating state in a good agreement with the DMFT,
and the quasiparticle weight $Z$ obtained in the large $U$ limit reproduces the famous Gutzwiller approximation $g_t=2x/(1+x)$.\cite{gutzwiller1963,florian1990,hassan2010} 
Moreover, the $U(1)$ form of the slave-spin method proposed by Yu and Si\cite{yu2012} can obtain the correct non-interacting limit within the same framework.
Very recently, it has been shown that the Landau-Ginzburg theory can be constructed based on the slave-spin method,\cite{yu2017,komijani2017}
and it has been pointed out that the slave-spin method in the single-site approximation at the saddle point level is a subset of the DMFT.\cite{komijani2017}
 
Motivated by these developments in the slave-spin method, in this paper we formulate the cluster slave-spin method to systematically improve the accuracy of this method to solve the Hubbard model 
with or without broken symmetry.
As a demonstration, the AFM state in the single band Hubbard model is studied using the cluster slave-spin method with the two and four-site clusters.
While the Hubbard gap, which is the charge gap due to the doubly-occupied states, scales with the Hubbard interaction $U$ as expected, the AFM gap $\Delta$,
which is the gap in the spinon dispersion in the AFM state, exhibits the standard mean-field behavior of $\Delta\sim U$ in the weak coupling limit and the 
superexchange behavior of $\Delta \sim t^2/U$ in the strong coupling limit. 
The holon-doublon correlator as functions of $U$ and doping $x$ is also computed, which exhibits a strong tendency toward the holon-doublon binding in the strong coupling regime.
In addition, we find that the quasi-particle weight in the paramagnetic state is also systematically improved beyond the Gutzwiller approximation as the cluster is enlarged from a 
single site to 4 sites.
Our results demonstrate that the cluster slave-spin method is a promising tool to study the strongly correlated system.

\section{Formalism}
We start from a general multiorbital Hubbard model of
\beq
H&=&H_t + H_U\nn\\
H_t&=& \sum_{i,j,\alpha,\beta,\sigma} \big(-t^{\alpha\beta}_{ij} - \Delta_\alpha \delta_{\alpha,\beta}\big)d^\dagger_{i\alpha\sigma} d_{j\beta\sigma},
\eeq
where $d^\dagger_{i\alpha\sigma}$ creates an electron with physical spin $\sigma$ on the orbital $\alpha$ at site $i$, $\Delta_\alpha$ is the crystal field splitting on the orbital $\alpha$, 
and $H_U$ represents the multiorbital Hubbard interactions.
In the slave-spin method, the electron creation operator is rewritten as
\be
d^\dagger_{i\alpha\sigma}\equiv \hat{O}_{i\alpha\sigma} f^\dagger_{i\alpha\sigma},
\label{ec}
\ee
where $\hat{O}_{i\alpha\sigma}$ is a spin operator associated an auxiliary $1/2$ spin representing the charge fluctuations of the electron with the physical spin $\sigma$ at site $i$, 
and the physical spin excitations are described by the fermionic spinon $f_{i\alpha\sigma}$. 
Note that since a slave spin is introduced for each physical spin $\sigma$, all the operators asscoiated with the slave spins has an index of $\sigma$ as well. 
In the original $Z_2$ slave-spin method, de Medici {\it et. al.} proposed\cite{demedici2005,demedici2009,hassan2010}
\be
\hat{O}_{i\alpha\sigma} = S^+_{i\alpha\sigma} + c_{i\alpha\sigma} S^-_{i\alpha\sigma},
\ee
where $c_{i\alpha\sigma}$ is a complex number used to ensure the correct Gutzwiller result of the quasiparticle weight $Z=2x/(1+x)$ at finite doping $x$ in the 
large $U$ limit.
The Hilbert space in the slave-spin formalism is enlarged, and we can in principle project out the unphysical states by enforcing the following constraint:
\be
S^z_{i\alpha\sigma} = f^\dagger_{i\alpha\sigma}f_{i\alpha\sigma} - \frac{1}{2}
\label{constraint}
\ee
The above constraint can be treated by introducing the Lagrangian multiplier $\lambda_{i\alpha\sigma}$, and at the level of saddle point approximation,
we find the solution with the constraint being satisfied on the average by making the assumption of $\lambda_{i\alpha\sigma} = \lambda_{\alpha\sigma}$.

Although it has been shown that this $Z_2$ slave-spin method can reproduce the featureless Mott insulating state in good agreements with the dynamical mean-field theory (DMFT) even at the level
of saddle point approximation, it fails to reproduce the non-interacting limit in which the quasiparticle weight $Z$ should be just 1 and the band structures should remain unchanged as 
$U=0$.\cite{yu2012}
The reason for this failure is that the Lagrangian multiplier $\lambda_{\alpha\sigma}$ has to be non-zero to yield $Z=1$ even as $U=0$, and a non-zero Lagrangian multiplier can result 
in an shift of the chemical potential in the spinon Hamiltonian.
For a multiorbital system, such a chemical potential shift is generally orbital-dependent, which modifies the band structures even without the interactions.
In other words, it is impossible for the $Z_2$ slave-spin formalism to obtain simultaneously $Z=1$ and the unchanged band structures in the non-interacting limit.

To fix this problem, Yu and Si proposed the $U(1)$ slave-spind method\cite{yu2012} in which the spin operator $\hat{O}_{i\alpha\sigma}$ in Eq. \ref{ec} is chosen to be
\be
\hat{O}_{i\alpha\sigma} = S^+_{i\alpha\sigma}.
\ee
The slave-spin method with this choice is invariant under a $U(1)$ gauge transformation of $f^\dagger_{i\alpha\sigma} \to f^\dagger_{i\alpha\sigma} e^{-i\theta_{i\alpha\sigma}}$
and $S^+_{i\alpha\sigma} \to S^+_{i\alpha\sigma} e^{i\theta_{i\alpha\sigma}}$.
The Gutzwiller limit can be ensured using the Kotliar-Ruckenstein slave-boson mean-field theory in which the slave-spin is represented by the 'dressed' Schwinger bosons 
$\{a_{i\alpha\sigma},b_{i\alpha\sigma}\}$ as
\beq
S^+_{i\alpha\sigma} &=& P^+_{i\alpha\sigma} a^\dagger_{i\alpha\sigma} b_{i\alpha\sigma} P^-_{i\alpha\sigma},\nn\\
S^z_{i\alpha\sigma} &=& \frac{1}{2}\big(a^\dagger_{i\alpha\sigma} a_{i\alpha\sigma} - b^\dagger_{i\alpha\sigma} b_{i\alpha\sigma}\big),\nn\\
P^\pm_{i\alpha\sigma}&=&\frac{1}{\sqrt{1/2 \pm S^z_{i\alpha\sigma}}},
\label{sbosons}
\eeq 
with the constraint of 
\be
a^\dagger_{i\alpha\sigma} a_{i\alpha\sigma} + b^\dagger_{i\alpha\sigma} b_{i\alpha\sigma} = 1.
\label{sbosonscon}
\ee
The constraint to project out the unphysical Hilbert space is still the same one given in Eq. \ref{constraint}.
Finally, Yu and Si showed that an extra orbital-dependent chemical potential in the spinon Hamiltonian can be generated to guarantee the correct behavior in the non-interacting limit, and 
the $U(1)$ theory obtains the same results by the $Z_2$ theory as the interaction is turned on.
In other words, the $U(1)$ slave-spin method can capture both the non-interacting and strong coupling (Gutzwiller) limits correctly, which makes it a powerful tool to study the strongly correlated 
system.

\begin{figure}
\includegraphics[width=3.5in]{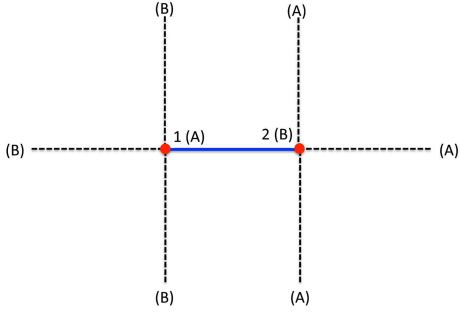}
\caption{\label{fig:2-site} Schematical illustration of the two-site cluster.}
\end{figure}

In light of the success of the $U(1)$ slave-spin method, we now extend the $U(1)$ slave-spin method to study symmetry-broken density wave states with a cluster slave-spin approach.
Generally speaking, in the density wave states the interactions produce an effective site-dependent chemical potential which is usually overestimated in the traditional mean-field theory in the 
strong coupling regime. The $U(1)$ slave-spin method can obtain the effective chemical potential self-consistently with the quantum fluctuations of the slave spins taken into account, 
if we generalize the single-site approach to include more sites in a cluster.
As a demonstration, we investigate the AFM state with the wavevector $\vec{Q}=(\pi,\pi)$ in the single band Hubbard model with only the nearest-neighbor hopping $t$, 
which requires a cluster of the slave spins. 
The smallest cluster is a two-site one shown in Fig. \ref{fig:2-site}, and the hopping Hamiltonian can be expressed as
\beq
H^{hop}_t &=& \sum_{\langle i,j\rangle,\sigma} -t S^+_{i\sigma}S^-_{j\sigma} f^\dagger_{i\sigma} f_{j\sigma} + H.c.\nn\\
&\approx& H^{hop,f}_t + H^{hop,S}_t,
\eeq
where $\langle i,j\rangle$ refers to the nearest neighbor, and we have adopted the saddle point approximation to decouple the original Hamiltonian into 
mean-field Hamiltonians for the slave spin ($H^{hop,f}_t$) and the spinon ($H^{hop,S}_t$) as
\beq
H^{hop,f}_t&=& \sum_{\langle i,j\rangle,\sigma} -t\langle S^+_{i\sigma}S^-_{j\sigma}\rangle f^\dagger_{i\sigma} f_{j\sigma} + H.c.,\nn\\
H^{hop,S}_t&=& \sum_{\langle i,j\rangle,\sigma} -t\langle f^\dagger_{i\sigma} f_{j\sigma}\rangle S^+_{i\sigma}S^-_{j\sigma} + H.c.
\eeq
In the single-site slave-spin method, it is assumed that the slave-spin degree of freedom is site-independent, thus we can make the approximations of 
$\langle S^+_{i\sigma}S^-_{j\sigma}\rangle \approx \vert \langle S^+_{\sigma}\rangle\vert^2$ in $H^{hop,f}_t$ and 
$S^+_{i\sigma}S^-_{j\sigma} \approx \langle S^+_{\sigma}\rangle S^-_{\sigma} + H.c.$ in $H^{hop,S}_t$.
In the two-site cluster, we need to include the difference between the sublattices $A$ and $B$, and $H^{hop,f}_t$ can be reduced to
\beq
H^{hop,f}_t&\approx& \sum_{\langle i,j\rangle,\sigma} -t\langle S^+_{A\sigma}\rangle\langle S^-_{B\sigma}\rangle f^\dagger_{i\sigma} f_{j\sigma} + H.c..\nn\\
\eeq
In order to inculde the quantum fluctuations enabled by the cluster, we have to consider more terms in $H^{hop,S}_t$ which are selected according to the following rules.
For the bond inside the cluster (the blue solid line in Fig. \ref{fig:2-site}) we can solve it exactly, and consequently we do not need to make further assumptions.
For the bond outside the cluster (the black dashed lines in Fig. \ref{fig:2-site}), we make the standard saddle-point decoupling. 
As a result, we have
\beq
H^{hop,S}_t&\approx& \sum_\sigma \epsilon^x_\sigma \big(S^+_{A\sigma}S^-_{B\sigma} + S^-_{A\sigma}S^+_{B\sigma}\big)\nn\\
&+& \big(\epsilon^x_\sigma + 2 \epsilon^y_\sigma\big) \big[\langle S^+_{A\sigma}\rangle S^-_{B\sigma} + \langle S^-_{B\sigma}\rangle S^+_{A\sigma} + H.c.\big],\nn\\
\eeq
where 
\be
\epsilon^{x(y)}_\sigma = -t\langle f^\dagger_{i\sigma} f_{i\pm\hat{x}(\hat{y}),\sigma}\rangle.
\label{spinonke}
\ee
Now we follow the recipe by Yu and Si\cite{yu2012} to expand 
\beq
S^-_{I\sigma}&\approx& \tilde{z}_{I\sigma} + \cdots\,\,\,,\,\,\,S^+_{I\sigma}\approx \tilde{z}_{I\sigma}^\dagger + \cdots,\nn\\
\tilde{z}_{I\sigma}&=& \frac{b^\dagger_{I\sigma} a_{I\sigma}}{\sqrt{(1/2)^2 - \langle S^z_{I\sigma}\rangle^2}},
\label{dresssb}
\eeq
and move the extra terms into the spinon Hamiltonian via the constraint. 
We then arrive at the mean-field Hamiltonians for the spinon and the slave spin as
\beq
H^{MF,f} &=&\sum_{\langle i,j\rangle,\sigma} \big[-t Z - \delta_{i,j} \big(\mu + \lambda_{I\sigma} - \tilde{\mu}_{I\sigma}\big) 
\big] f^\dagger_{i\sigma} f_{j\sigma}+ H.c.,\nn\\
H^{MF,S} &=&  \sum_\sigma \lambda_{A\sigma} S^z_{A\sigma} + \lambda_{B\sigma} S^z_{B\sigma} + 
\epsilon^x_\sigma \big(\tilde{z}^\dagger_{A\sigma}\tilde{z}_{B\sigma} + H.c.\big)\nn\\
&+& \big(\epsilon^x_\sigma + 2 \epsilon^y_\sigma\big) \big[\langle \tilde{z}^\dagger_{A\sigma}\rangle \tilde{z}_{B\sigma} + \langle \tilde{z}^\dagger_{B\sigma}\rangle \tilde{z}_{A\sigma} 
+ H.c.\big]\nn\\
&+& H^U,
\eeq
where
\be
\tilde{\mu}_{I\sigma} = \frac{4Z\langle S^z_{I\sigma}\rangle \big(\epsilon^x_\sigma + \epsilon^y_\sigma\big)}{(\frac{1}{2})^2 - \langle S^z_{I\sigma}\rangle^2}
\label{tildemu}
\ee
is the effective chemical potential shift generated by the qutanum fluctuations from the slave spins,
$Z=\langle \tilde{z}^\dagger_{A\sigma}\rangle\langle \tilde{z}_{B\sigma}\rangle$ is the quasiparticle weight,
$I = A (B)$ if $i\in$ sublattice $A (B)$,
$\lambda_{I\sigma}$ is the Lagrangian multiplier to satisfy the constraint in Eq.\ref{constraint} on the average,
and $H^U$ is the Hubbard interaction which can be written in terms of the slave spins as
\be
H^U = U\sum_{I=A,B} \big[(S^z_{I\uparrow}+\frac{1}{2})(S^z_{I\downarrow}+\frac{1}{2})\big].
\ee
Finally, we can self-consistently compute $\big(\langle \tilde{z}_{I\sigma}\rangle,\lambda_{I\sigma}\big)$, and the quasiparticle weight $Z$ can be obtained.

It is worth mentioning that $H^{MF,f}$ can be Fourier transformed into the $\vec{k}$ space as
\beq
H^{f,MF} &=& \sum_{\vec{k},\sigma} \xi(\vec{k}) f^\dagger_{\vec{k}\sigma}f_{\vec{k}\sigma} + \Delta_\sigma f^\dagger_{\vec{k}+\vec{Q}\sigma} f_{\vec{k}\sigma},\nn\\
\xi(\vec{k})&=& -2t Z(\cos k_x + \cos k_y) - \mu_{eff},\nn\\
\mu_{eff}&=&\mu - \frac{1}{2}\big(\tilde{\mu}_{A\sigma}-\lambda_{A\sigma} + \tilde{\mu}_{B\sigma}-\lambda_{B\sigma}\big),\nn\\
\Delta_\sigma&=&\frac{1}{2}\big(\tilde{\mu}_{A\sigma}-\lambda_{A\sigma} - \tilde{\mu}_{B\sigma}+\lambda_{B\sigma}\big).
\label{afmspinon}
\eeq
Although Eq. \ref{afmspinon} bears the same form of the traditional mean-field theory for the AFM state, there are two important differences.
Firstly, the hopping parameter $t$ is now renormalized by $Z$, which takes into account the reduction of the bandwidth due to the Hubbard interaction. 
Secondly, the AFM gap $\Delta=\Delta_\uparrow=\Delta_\downarrow$ is determined by differences in $\lambda_{I\sigma}$ and $\tilde{\mu}_{I\sigma}$ between two sublattices,
which are self-consistently computed with the quantum fluctuations of the slave spins taken into account.

\begin{figure}
\includegraphics[width=3.5in]{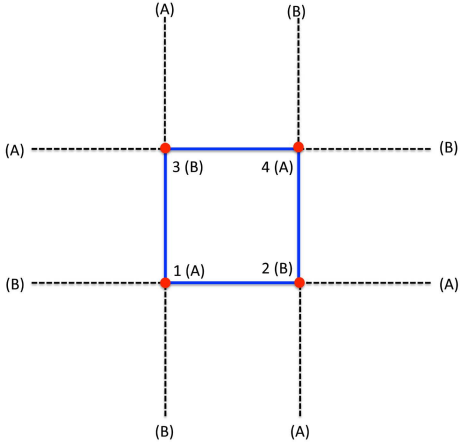}
\caption{\label{fig:4-site} Schematical illustration of the four-site cluster.}
\end{figure}

To see whether our formlalism indeed includes more quantum fluctuations as the number of sites in the cluster increases, we will perform the same calculations with a four-site cluster as shown in 
Fig. \ref{fig:4-site}. The mean-field Hamiltonians and equations for the four-site cluster can be derived in the same way. 
The spinon Hamiltonian remains the same, but the slave-spin Hamiltonian reads
\beq
&&H^{MF,S}_{4site} =  H^U_{4site} + \sum_{I=1}^4\sum_\sigma \lambda_{I\sigma} S^z_{I\sigma}\nn\\
&+&\sum_\sigma\big\{ \epsilon^x_\sigma \big(\tilde{z}^\dagger_{1\sigma}\tilde{z}_{2\sigma} + \tilde{z}^\dagger_{3\sigma}\tilde{z}_{4\sigma}\big)
+\epsilon^y_\sigma \big(\tilde{z}^\dagger_{1\sigma}\tilde{z}_{3\sigma} + \tilde{z}^\dagger_{2\sigma}\tilde{z}_{4\sigma}\big)\nn\\
&+&\epsilon^x_\sigma \big(\langle \tilde{z}^\dagger_{1\sigma}\rangle \tilde{z}_{2\sigma} + \langle \tilde{z}_{2\sigma}\rangle \tilde{z}^\dagger_{1\sigma} + 
\langle \tilde{z}^\dagger_{3\sigma}\rangle \tilde{z}_{4\sigma} + \langle \tilde{z}_{4\sigma}\rangle \tilde{z}^\dagger_{3\sigma}\big)\nn\\
&+&\epsilon^y_\sigma \big(\langle \tilde{z}^\dagger_{1\sigma}\rangle \tilde{z}_{3\sigma} + \langle \tilde{z}_{3\sigma}\rangle \tilde{z}^\dagger_{1\sigma} +
\langle \tilde{z}^\dagger_{2\sigma}\rangle \tilde{z}_{4\sigma} + \langle \tilde{z}_{4\sigma}\rangle \tilde{z}^\dagger_{2\sigma}\big) + H.c.\big\},\nn\\
&&H^U_{4site} = U\sum_{I=1}^4 \big[(S^z_{I\uparrow}+\frac{1}{2})(S^z_{I\downarrow}+\frac{1}{2})\big],
\eeq
where sites 1 and 4 belong to the sublattice $A$ and sites 2 and 3 belong to the sublattice $B$.
It can be easily seen that the dimension of the matrix corresponding to the slave-spin Hamiltonian is $4^N$, where $N$ is the number of sites in the cluster.
As a result, the more sites in the cluster there are, the more inter-site fluctuations will be included. 
The inclusion of the inter-site fluctuations appropriately is critically important to describe the AFM state in the strong-coupling limit.
In this paper, all the results presented are  at zero temperature.
 
\section{Results}
\begin{figure}
\includegraphics[width=3in]{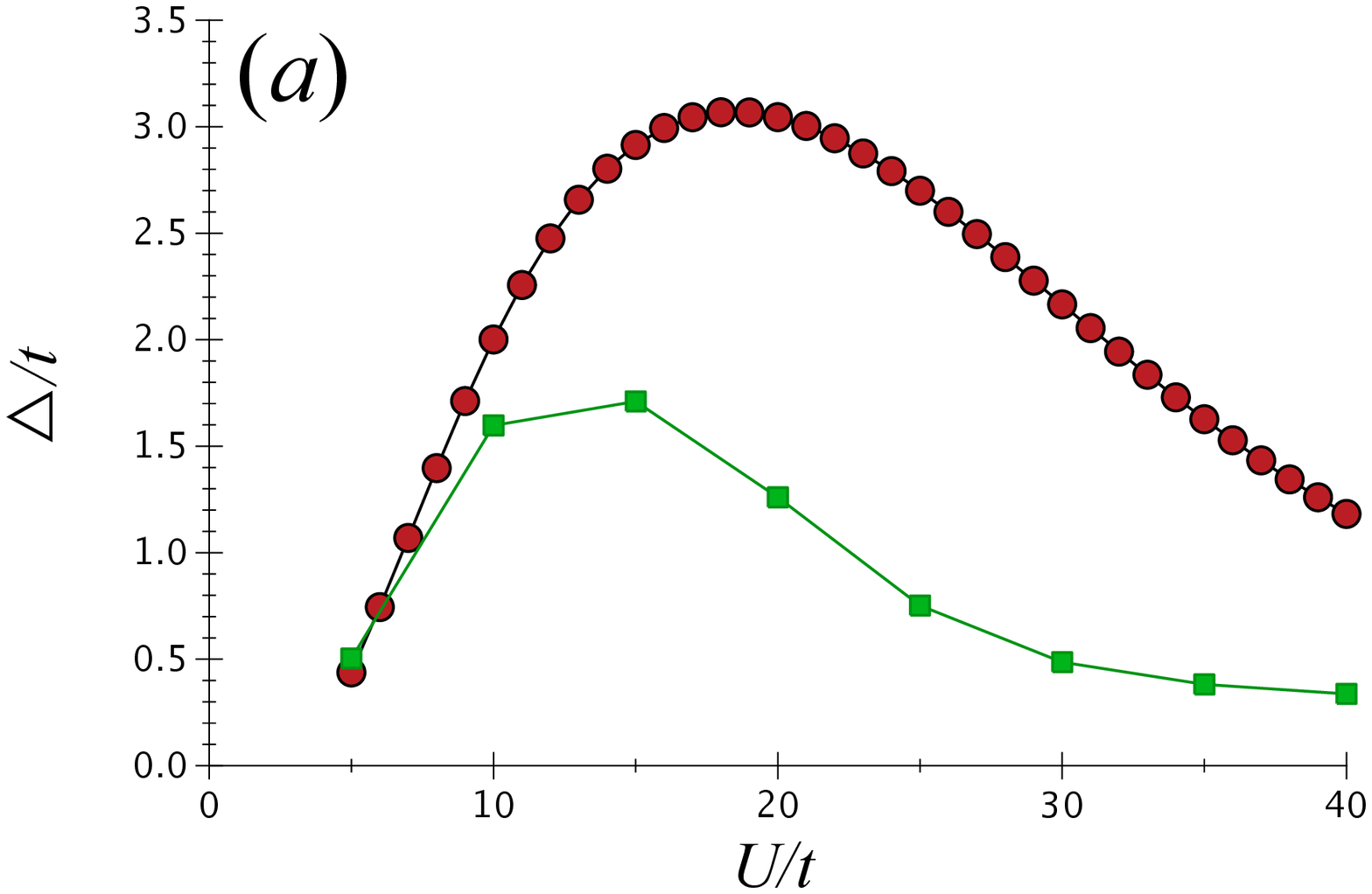}
\includegraphics[width=3in]{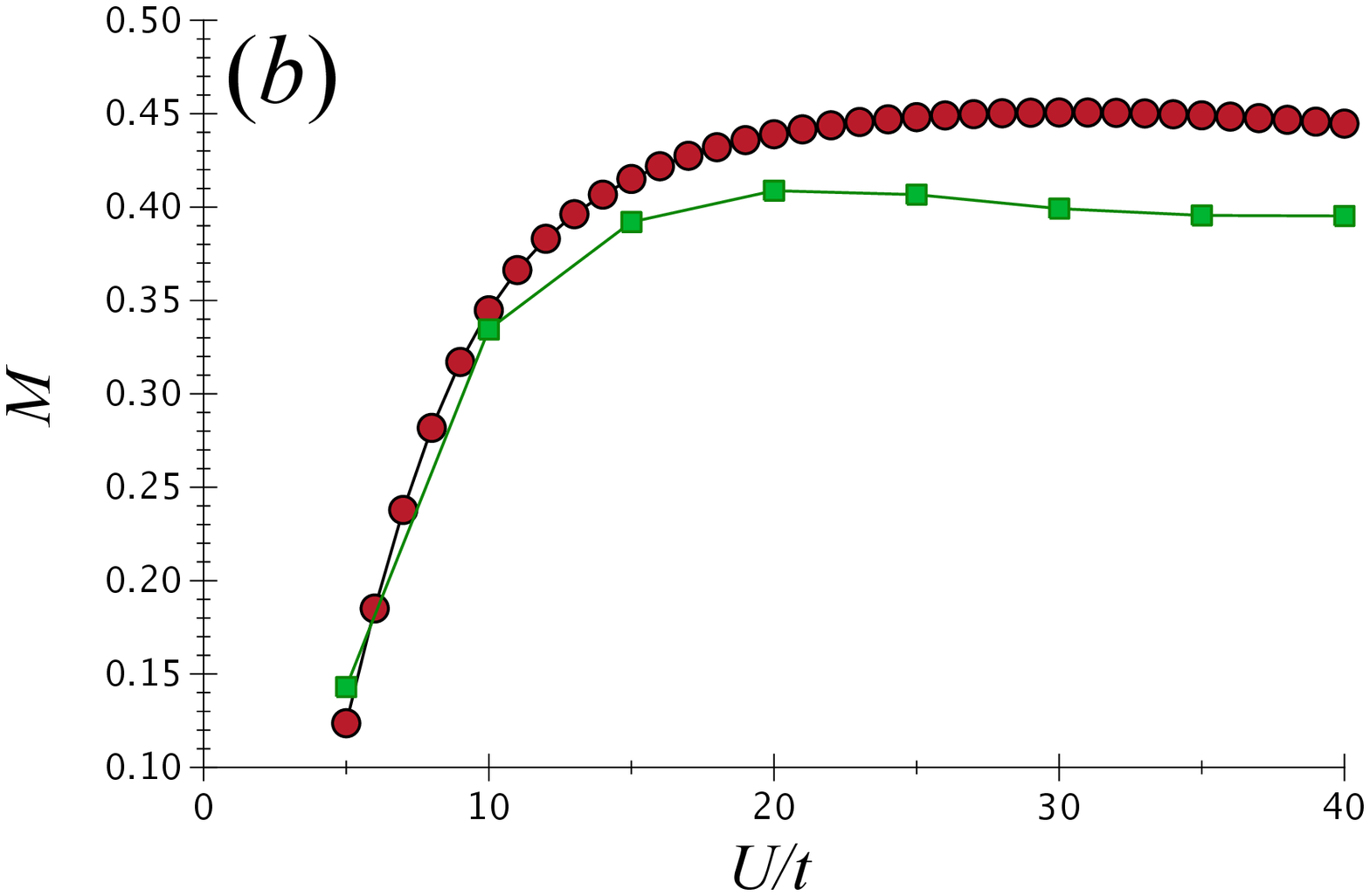}
\caption{\label{fig:gapm-x002} (Color online) (a) The AFM gap $\Delta$ and (b) the magnetic moment $M$ as a function of $U$ at $x=0.02$ computed by the two-site cluster 
(red circle) and the four site cluster (green square).}
\end{figure}

The evolution of the AFM gap $\Delta$ obtained by our cluster slave-spin method as a function of $U$ at the doping level $x=0.02$ is plotted in Fig. \ref{fig:gapm-x002}(a).
At small $U$, our results reproduces the traditional mean-field behavior $\Delta\sim U$ very well.
Moreover, we see that the results obtained by the two and four-site clusters are similar at very small $U$, indicating that the inter-site fluctuations are not important and consequently 
the traditional mean-field picture is valid. 
As $U$ increases, we observe that $\Delta$ reaches a maximum around a critical value $U_c$ and then decreases as $\Delta\sim t^2/U$ at large $U$, which is 
consistent with the recent variational Monte Carlo results.\cite{wuhk2017}
This crossover  signals that at the large $U$ limit, the mechansim for the AFM state becomes the superexchange mechanism, which is exactly the physics overlooked in the traditional mean-field theory.
We emphasize that such a crossover can never be obtained without taking into account the local charge fluctuations appropriately.
The cluster slave-spin method is designed to treat the charge fluctuations in $H^{S,MF}$ correctly in both the non-interacting and the strong coupling limits, 
thus it can naturally obtain the crossover within a single framework.

Fig. \ref{fig:gapm-x002}(b) presents the magnetic moment $M$ which is defined in the usual way as
\be
M = \frac{1}{2 \Omega}\sum_i e^{i\vec{Q}\cdot \vec{R}_i} \big(\langle f^\dagger_{i\uparrow} f_{i\uparrow}\rangle - \langle f^\dagger_{i\downarrow} f_{i\downarrow}\rangle\big).
\ee
At small $U$, we find the traditional mean-field behavior of $M\sim U$ in both clusters as expected.
As $U$ increases, we observe that the magnetic moment saturates at $M\approx 0.45$ in the two-site cluster and at $M\approx 0.4$ in the four-site cluster.
Note that the traditional mean-field theory predicts $M\sim (1-x)/2$ (0.49 at $x=0.02$) in the large $U$ limit, and the quantum Monte Carlo calculation on the $t-J$ model at half-filling
obtains $M\approx 0.307$.\cite{sandvik1997}
Consequently, the fact of the smaller saturated magnetic moment obtained by the cluster slave-spin method
indicates its ability to capture the inter-site quantum fluctuations from the local Hubbard interaction.

\begin{figure}
\includegraphics[width=3in]{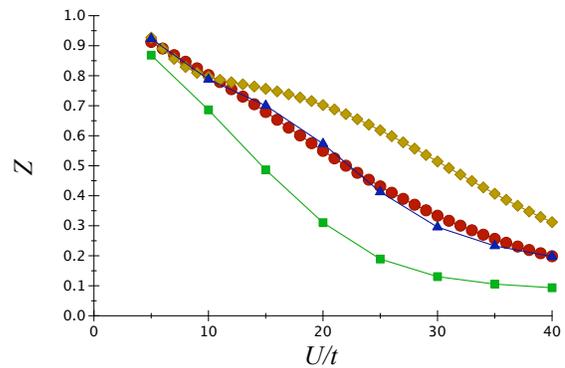}\caption{\label{fig:z-x002} (Color online) The quasiparticle weight $Z$ as a function of $U$ at $x=0.02$ obtained by the two and four-site clusters, and 
the generalized Gutzwiller approximation in AFM state given in Eq. \ref{gt} (yellow diamond for the two-site and blue triangle for the four site clusters).}
\end{figure}

Now we discuss the differences in the results obtained by the two and four-site clusters. For the AFM gap $\Delta$, we find that the critical value $U_c$ for the crossover decreases 
from $U_c/t\sim 18$ in the two-site cluster to $U_c/t\sim 14$ in the four-site cluster, and in the large $U$ regime $\Delta$ is smaller in the four-site cluster. 
To understand this trend, let's analyze the AFM gap $\Delta$ in Eq. \ref{afmspinon}. 
The main contribution to $\Delta$ is from the difference between sublattice $A$ and $B$ 
in the effective chemical potential generated by the slave-spin fluctuations ($\tilde{\mu}_{A\sigma} - \tilde{\mu}_{B\sigma}$), and from Eq. \ref{tildemu} we see that $\tilde{\mu}_{I\sigma}$ 
is proportional to the quasiparticle weight $Z$. In other words, the behavior of the quasiparticle weight $Z$ plays an important role in $\Delta$.

Before analyzing the quasiparticle weight $Z$, we introduce the generalized Gutzwiller approximation in the AFM states derived 
in Refs. [\onlinecite{ogawa1975}] and [\onlinecite{abram2013}], which can be expressed as
\beq
g_t =&& \frac{1-x-2 d}{1-x-2rw} \big(\sqrt{\frac{(1-w)(x+d)}{1-r}} + \sqrt{\frac{wd}{r}}\big)\nn\\
&\times&\big(\sqrt{\frac{(1-r)(x+d)}{1-w}} + \sqrt{\frac{rd}{w}}\big),
\label{gt}
\eeq
where $r$ ($w$) is the average electron occupation number per site with spin up (down) given as $r= \frac{1-x}{2} + M$ ($w= \frac{1-x}{2} - M$).
$d$ is the average density of the doubly-occupied site referred as the {\it doublon}. 
This term $g_t$ represents the renormalization of the kinetic hopping terms in the Gutzwiller approximation, which is just same as the quasiparticle weight $Z$.
Since the charge fluctuations are described by the slave spins, we should compute the doublon number within the slave-spin degrees of freedom.
In the slave-spin sector, the doublon state at site $i$ is the eigen state of $S^z_{i\sigma}$ such that 
\be
S^z_{j\sigma}\vert i,doublon\rangle = +\frac{\delta_{i,j}}{2}\vert i,doublon\rangle.
\ee
Consequently, the number operator of the doublon at site $i$ can be expressed as
\be
\hat{D}_i = \hat{n}^a_{i\uparrow} \hat{n}^a_{i\downarrow},
\label{dn}
\ee
where $\hat{n}^a_{i\sigma} = a^\dagger_{i\sigma} a_{i\sigma}$ is the number operator of the Schwinger bosons defined in Eq. \ref{sbosons}. 
Given that we limit our attention to solutions that are homogeneous in the charge degrees of freedom, we have $d=\langle \hat{D}_i\rangle$.
Fig. \ref{fig:z-x002} plots the quasiparticle weight $Z$ as a funtion of $U$ obtained by the two and four-site clusters and corresponding generalized Gutzwiller factors $g_t$.
We can see that in the small $U$ limit, the results obtained by both clusters are close to each other as expected.
In the large $U$ limit, however, our cluster slave-spin method usually obtains a smaller $Z$ compared to the generalized Gutzwiller factor $g_t$.
Furthermore, $Z$ obtained by the four-site cluster is generally smaller than that by the two-site cluster.
These results strongly suggest that as the size of the cluster increases, more inter-site quantum fluctuations are taken into account and consequently the results can be improved.

\begin{figure}
\includegraphics[width=3in]{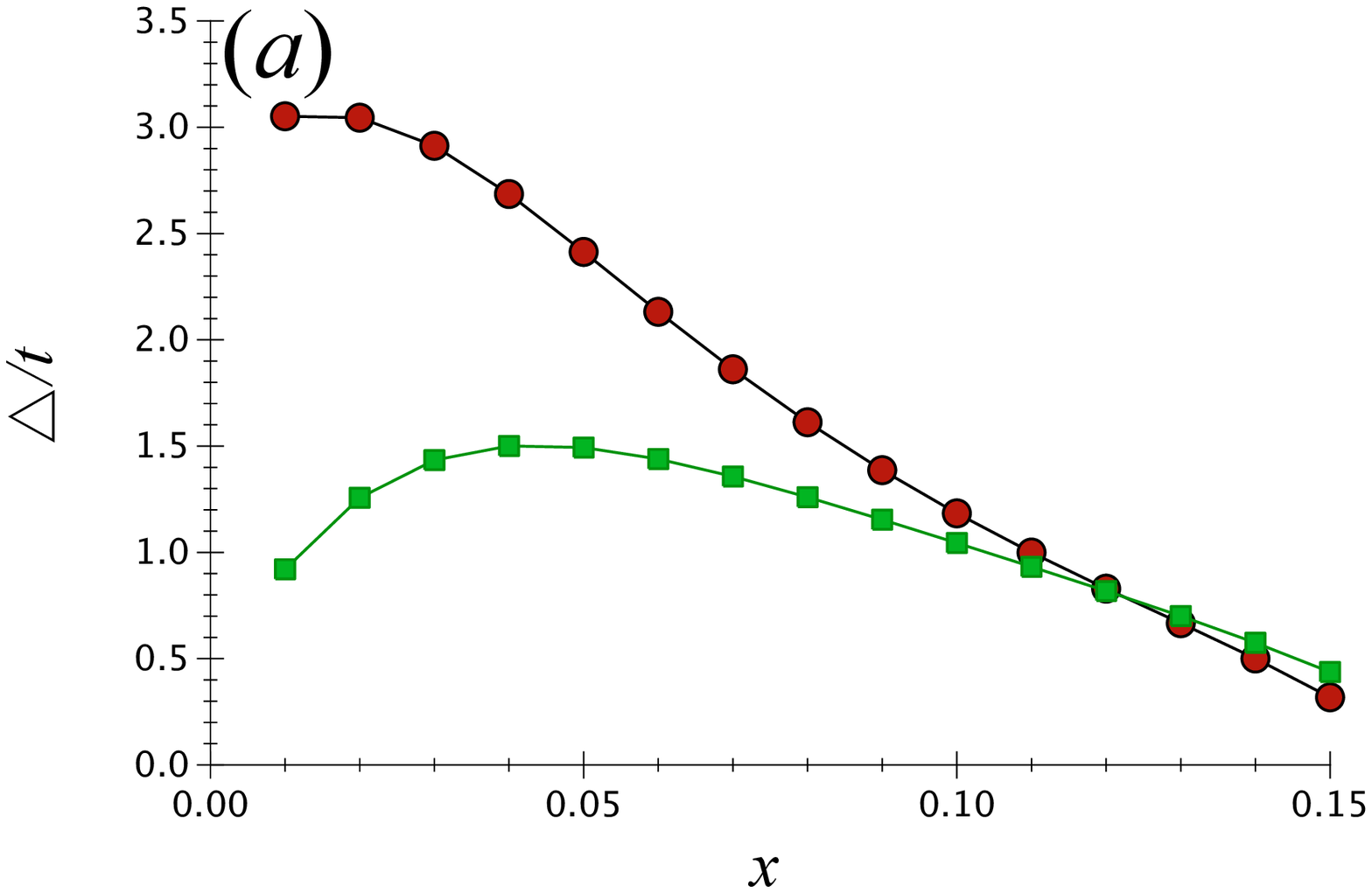}
\includegraphics[width=3in]{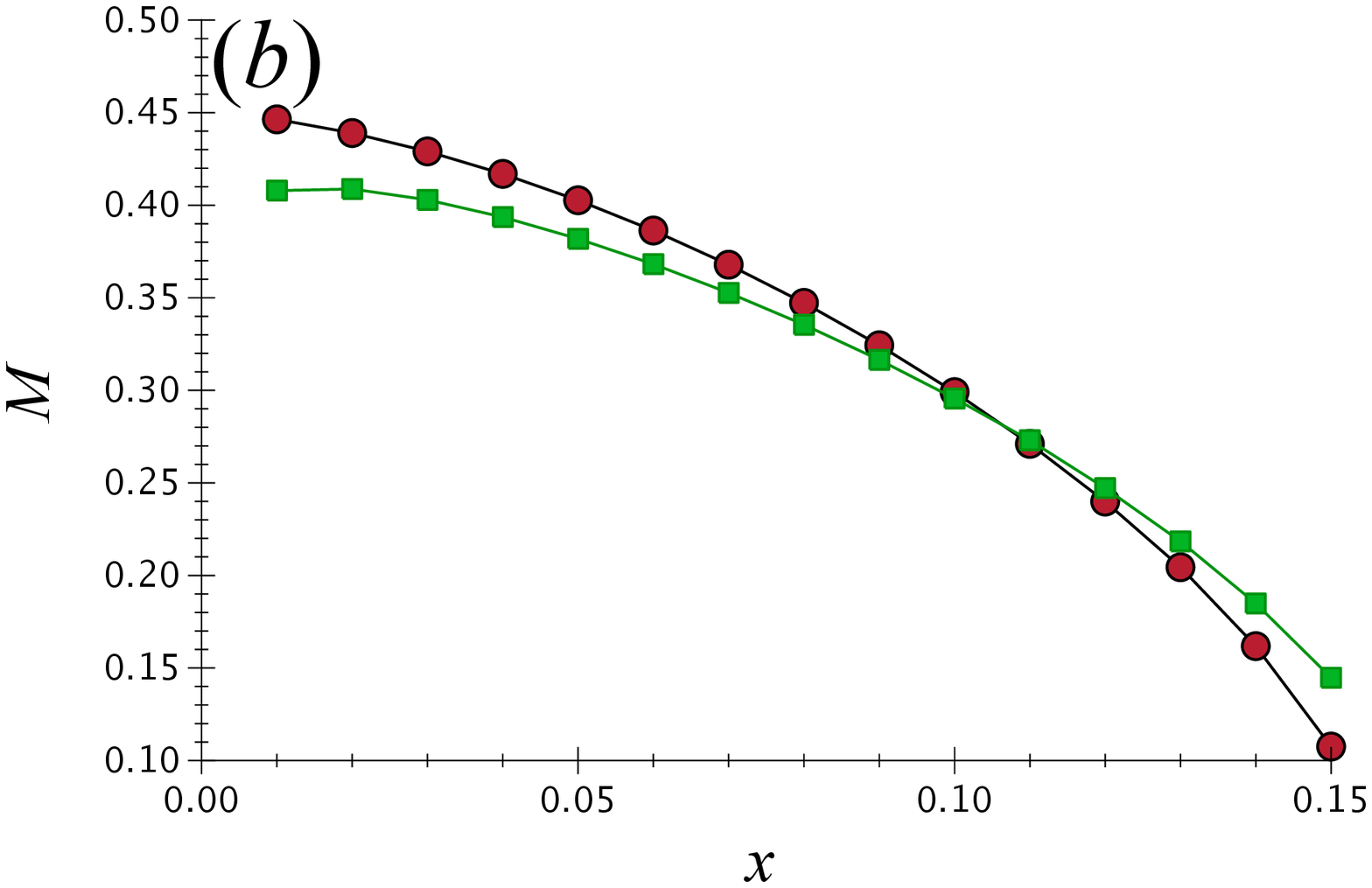}
\includegraphics[width=3in]{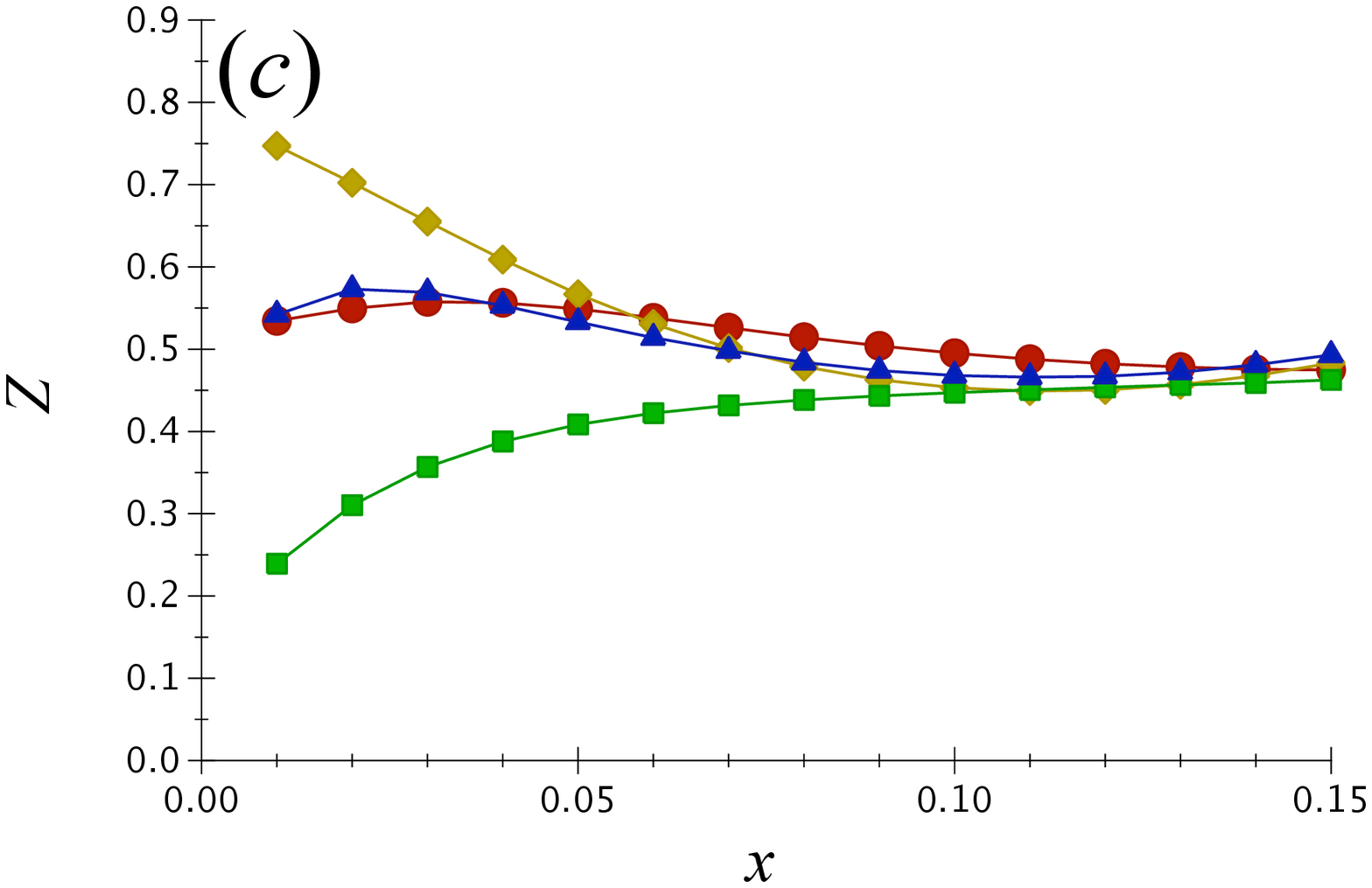}
\caption{\label{fig:gapmz-u20} (Color online) (a) The AFM gap $\Delta$ and (b) the magnetic moment $M$ as a function of the doping $x$ with $U/t=20$ 
computed by the two-site cluster (red circle) and the four site cluster (green square). (c) The quasiparticle weight $Z$ obtained by the the two-site, 
the four-site clusters, and the generalized Gutzwiller approximation in AFM state derived in Refs. [\onlinecite{ogawa1975}] and [\onlinecite{abram2013}] (yellow diamond for the 
two-site and blue triangle for the four-site clusters).}
\end{figure}

Finally, we present $\Delta$, $M$, and $Z$ as a function of the doping $x$ with $U/t=20>U_c/t$ in Fig. \ref{fig:gapmz-u20}.
We find that at large doping ($x>0.1$) both two-site and four-site clusters give similar results, which indicates that the inter-site correlation is less important at large doping
despite the fact that $Z$ is still small in this regime.
As a result, at large doping the system can be considered as a correlated Fermi liquid in which the bandwidth is strongly renormalized but a traditional mean-field picture is still valid.
On the other hand, at small doping both $Z$ and $\Delta$ changes dramatically as the size of the cluster increases from two to four, signaling that the strong coupling physics 
dominates in this regime and consequently the inter-site correlation is important. 
Moreover, it is straightfoward to find that $g_t$ at zero doping becomes
\beq
g_t(x=0) &=& \frac{(1-2 d)d}{1-2rw} \big(\sqrt{\frac{(1-w)}{1-r}} + \sqrt{\frac{w}{r}}\big)\nn\\
&\times&\big(\sqrt{\frac{(1-r)}{1-w}} + \sqrt{\frac{r}{w}}\big),\nn\\
&\propto& d,
\eeq
indicating that $g_t(x=0)$ approaches a constant as long as $d$ is non-zero. This result suggests that $g_t(x=0)$ will never be zero at finite $U$ and extra quantum fluctuations have to be taken 
into zccount to obtain the featureless Mott insulator at zero doping. 
It is interesting to see that $Z$ obtained by the four-site cluster drops much more significantly than $Z$ obtained by the two-site cluster and $g_t$, which 
is another evidence of the four-site cluster including more quantum fluctuations.
In short, the cluster slave-spin method can obtain correct physics in both weak and strong coupling limits, and in principle the results can be improved by increasing 
the size of the cluster of the slave spins.

\section{Holon-doublon correlation}
\begin{figure}
\includegraphics[width=3in]{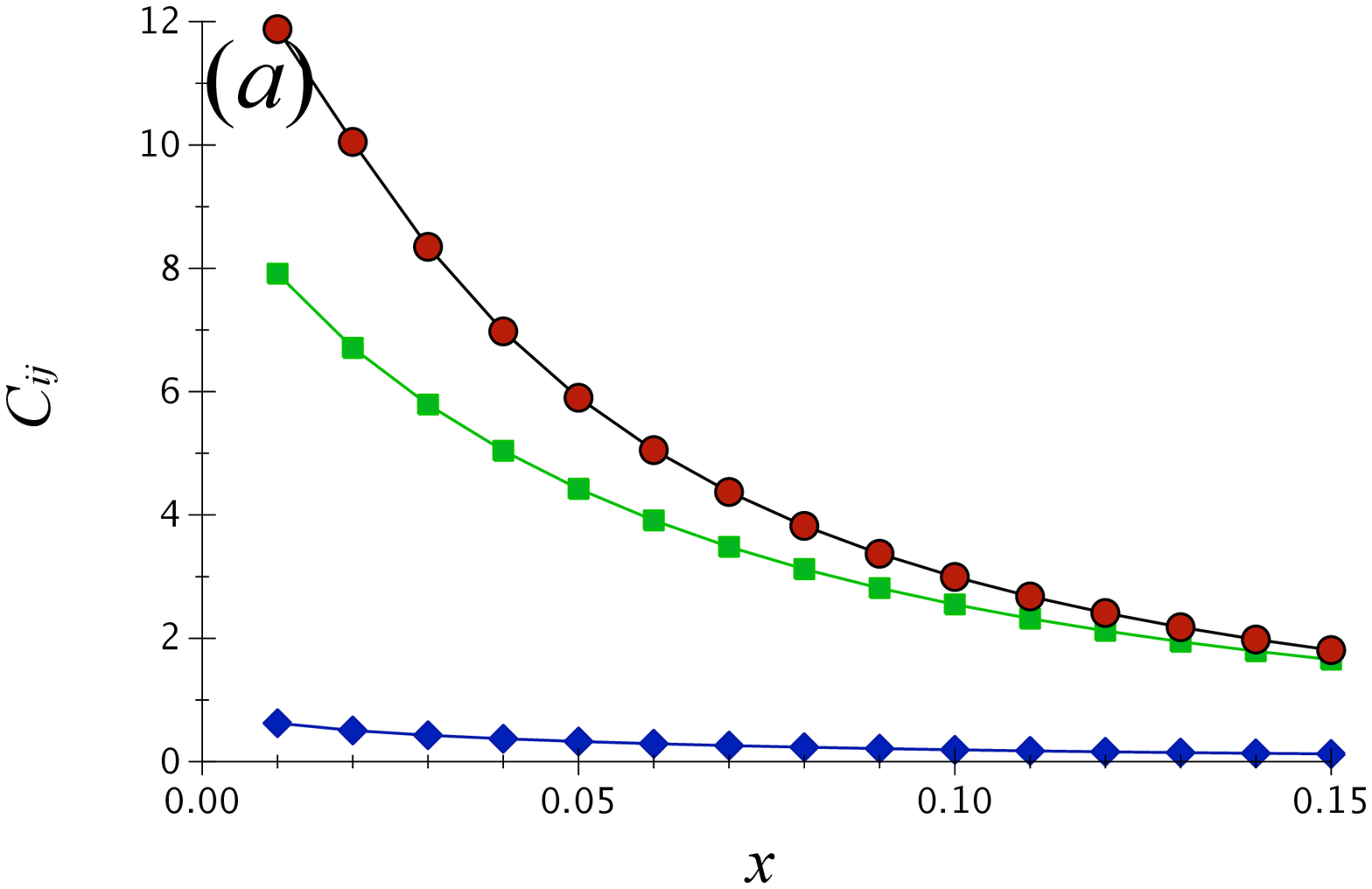}
\includegraphics[width=3in]{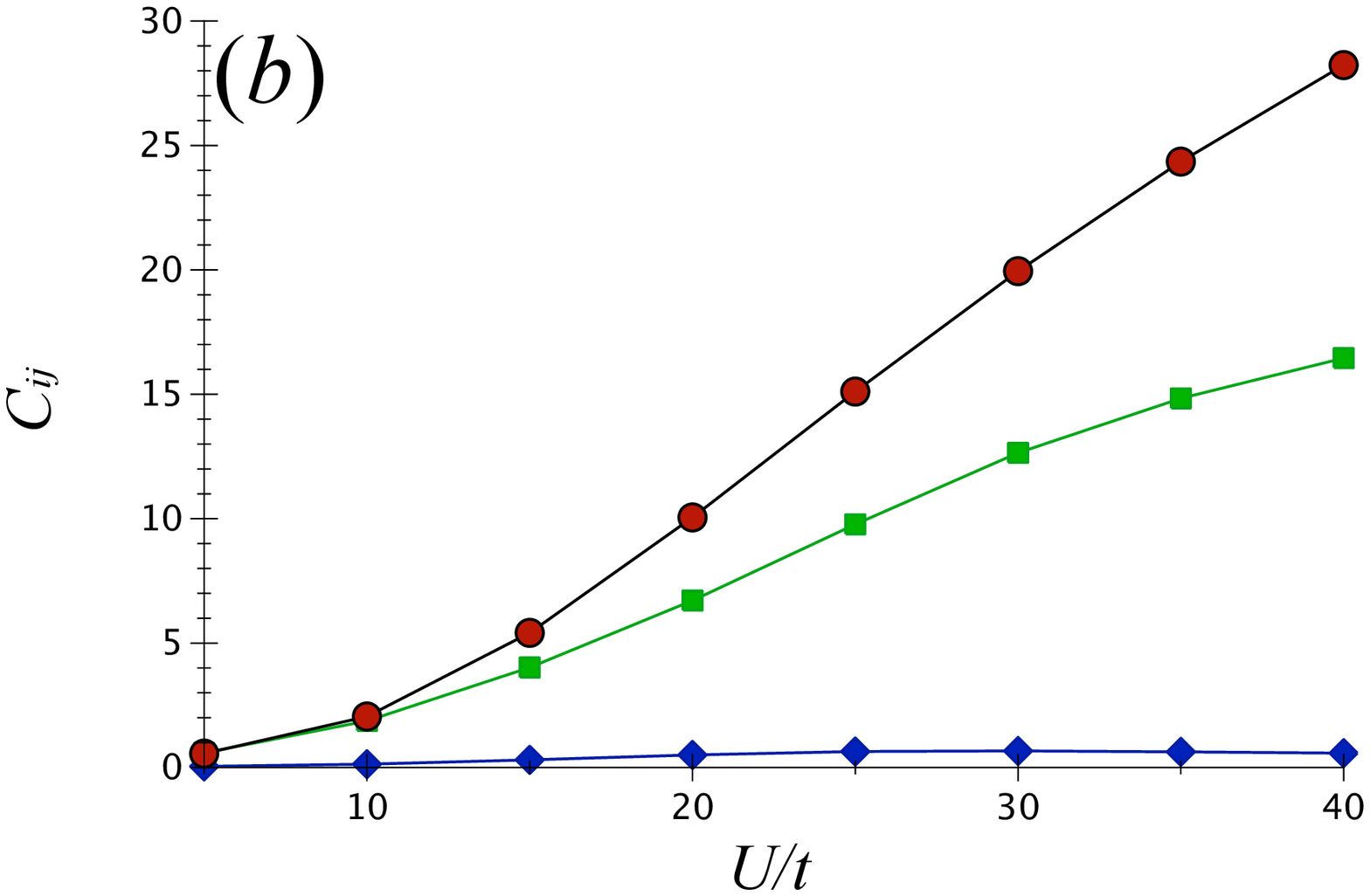}
\caption{\label{fig:cij-u20} (Color online) The holon-doublon correlation function within the four-site cluster for the nearest neighbor ($C_{12}$, green triangle) and the 
next nearest neighbor ($C_{14}$, blue diamond) (a) as a function of the doping 
$x$ with $U/t=20$ and (b) as a function of $U$ with $x=0.02$. As a comparison, $C_{12}$ (red circle) in the two-site cluster is also plotted.}
\end{figure}

To gain a better insight into the inter-site correlation, we calculate the holon-doublon correlation function for the four-site cluster. 
Following the same way to derive the doublon number operator in Eq. \ref{dn}, we obtain that the holon state at site $i$ is the eigen states of $S^z_{i\sigma}$ such that 
\be
S^z_{j\sigma}\vert i,holon\rangle = -\frac{\delta_{i,j}}{2}\vert i,holon\rangle,
\ee
and consequently the holon number operator ($\hat{N}_i$) at site $i$ can be expressed as
\be
\hat{N}_i = \hat{n}^b_{i\uparrow} \hat{n}^b_{i\downarrow},
\ee
where $\hat{n}^b_{i\sigma} = b^\dagger_{i\sigma} b_{i\sigma}$ is the number operator of the Schwinger bosons defined in Eq. \ref{sbosons}.
As a result, the holon-doublon correlation function can be evaluated by
\be
C_{ij} =\frac{\langle \hat{N}_i\hat{D}_j\rangle - \langle \hat{N}_i\rangle\langle\hat{D}_j\rangle }{\langle \hat{N}_i\rangle\langle\hat{D}_j\rangle}.
\ee
Physically, the holon and the doublon can not appear at the same site, which is automatically satisfied in our formalism since 
$C_{ii}$ is zero due to the constraint of the Schwinger bosons given in Eq. \ref{sbosonscon}.
Moreover, our formalism yields a relation between the average holon and doublon numbers of
\beq
\langle \hat{N}_i\rangle&=& \langle \hat{n}^b_{i\uparrow} \hat{n}^b_{i\downarrow}\rangle\nn\\
&=&\langle (1-\hat{n}^a_{i\uparrow})(1-\hat{n}^a_{i\downarrow})\rangle\nn\\
&=&1 - \big(\langle \hat{n}^a_{i\uparrow}\rangle + \langle\hat{n}^a_{i\downarrow}\rangle\big) + \langle \hat{D}_i\rangle\nn\\
&=& x + \langle \hat{D}_i\rangle,
\label{hd}
\eeq
which corresponds to the fact that the creation of a doublon must be accompanied by the creation of an additional holon due to the conservation of the electron number.

In the four-site cluster shown in Fig. \ref{fig:4-site} we can compute the nearest neighbor 
correlation function $C_{12}$ ($= C_{13}$ due to the $C_4$ symmetry) and the next nearest neighbor one $C_{14}$ which are shown in Fig. \ref{fig:cij-u20}(a).
At small doping, $C_{12}$ is much larger than $C_{14}$, indicating that the holon and doublon tend to be bound with each other. 
As the doping increases, $C_{12}$ decreases significantly and approaches $C_{14}$, suggesting that the holon-doublon binding disappears at large doping.
We further compute the holon-doublon correlation function at doping $x=0.02$ with different $U$, which is shown in Fig. \ref{fig:cij-u20}(b).
It can be seen clearly that the holon-doublon binding only appears at large $U$, which further proves that the holon-doublon binding is a universal feature in the strong coupling regime.
Note that in the two-site cluster because $C_{12}$ is the only holon-doublon correlation function that can be computed, there is no other phase space to place the holon and the doublon.
Therefore, the two-site cluster overestimates $C_{12}$ because it has less inter-site quantum fluctuations, which explains why $C_{12}$ is always smaller in the four-site cluster than in 
the two-site cluster.

\section{Quasiparicle weight and the holon-doublon correlation in the paramagnetic state}
To check whether the holon-doublon binding is truely a general feature of the Mott physics regardless the antiferromagnetism, in this section we study the paramagnetic state.
The Gutzwiller factor of the quasiparticle weight in the paramagnetic state can be derived by setting $M=0$ in Eq. \ref{gt}, which can be expressed as
\cite{ogawa1975,abram2013}
\be
g_t = \frac{1-x-2 d}{1-x-\frac{(1-x)^2}{2}} \big(\sqrt{x+d} + \sqrt{d}\big)^2.
\label{gtm0}
\ee
It is remarkable that the single-site $U(1)$ slave-spin method reproduces $g_t$ in Eq. \ref{gtm0} {\it exactly}, suggesting that the $U(1)$ slave-spin method has already reached the same 
level of accuracy as the Gutzwiller approximation even with the single slave-spin site. We provide the proof in the Appendix.
Below we will show that the results can be further improved with the larger size of the cluter.

\begin{figure}
\includegraphics[width=3in]{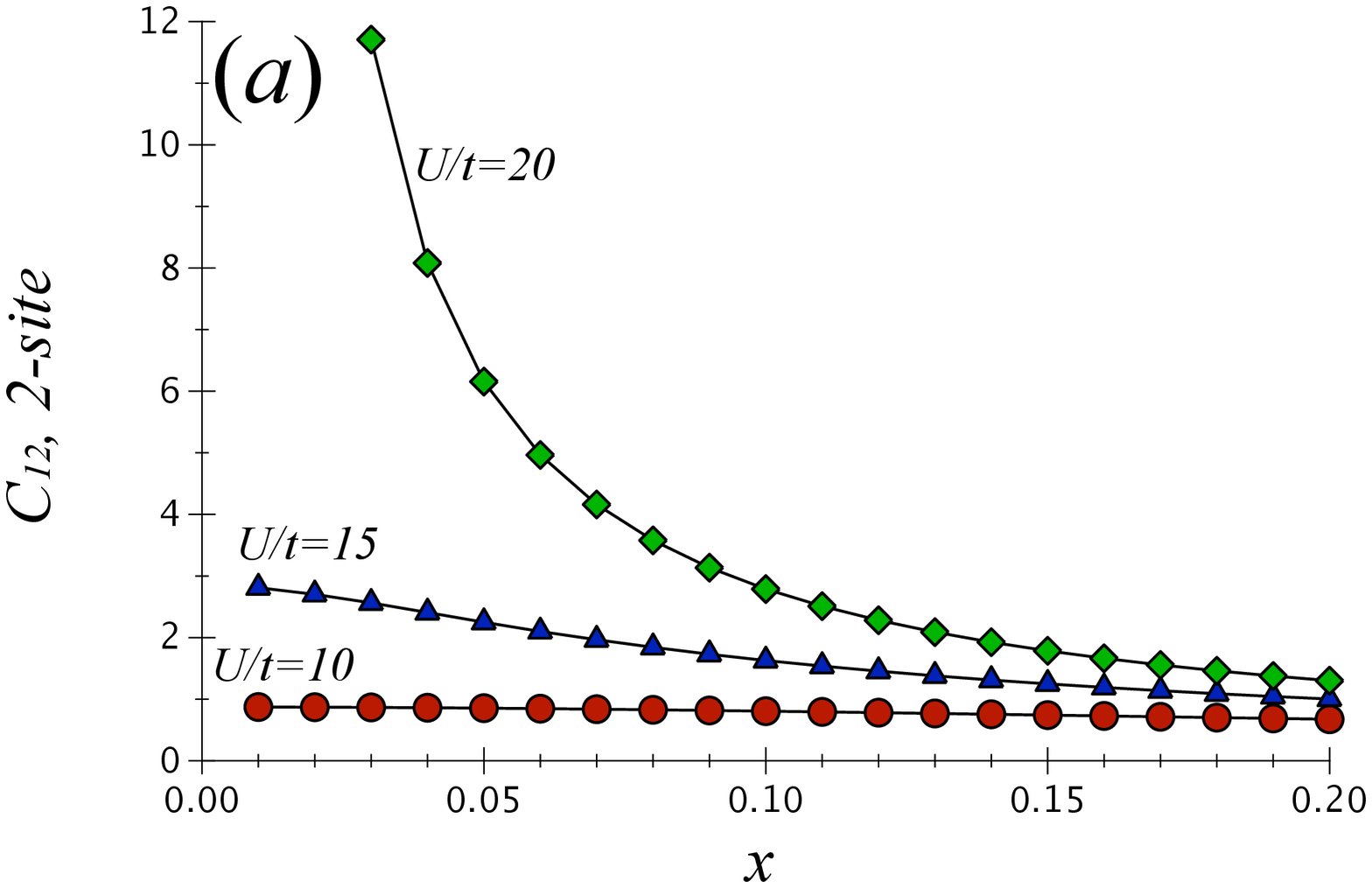}
\includegraphics[width=3in]{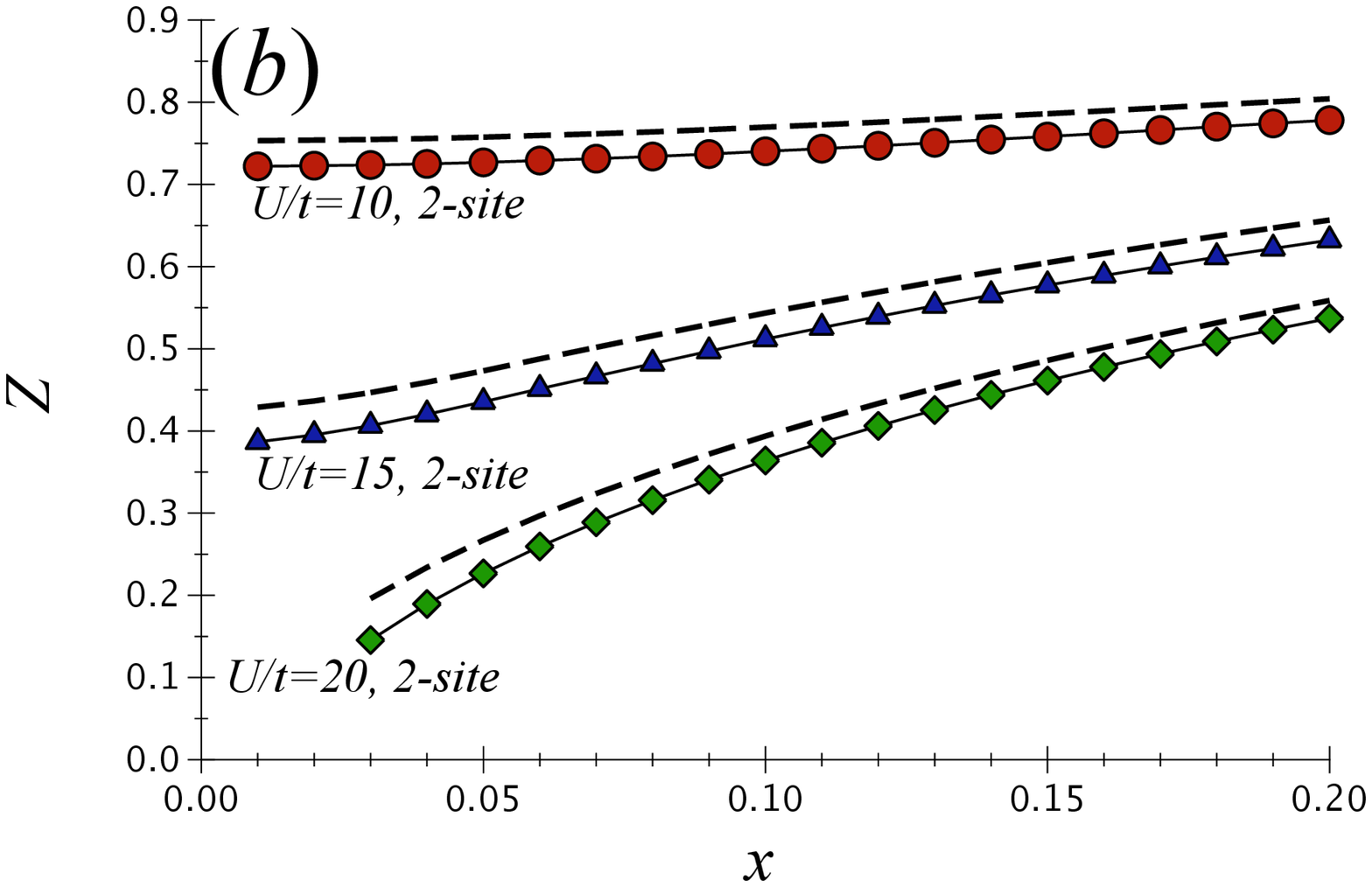}
\caption{\label{fig:para-2site} (Color online) (a) The holon-doublon correlation function for the nearest neighbor ($C_{12}$) and (b) the quasiparticle weight $Z$ as a function of doping $x$ at
$U/t=10,15,20$ in the paramagnetic state. The dashed lines represent the corresponding Gutzwiller factors $g_t$ for each case.}
\end{figure}

\begin{figure}
\includegraphics[width=3in]{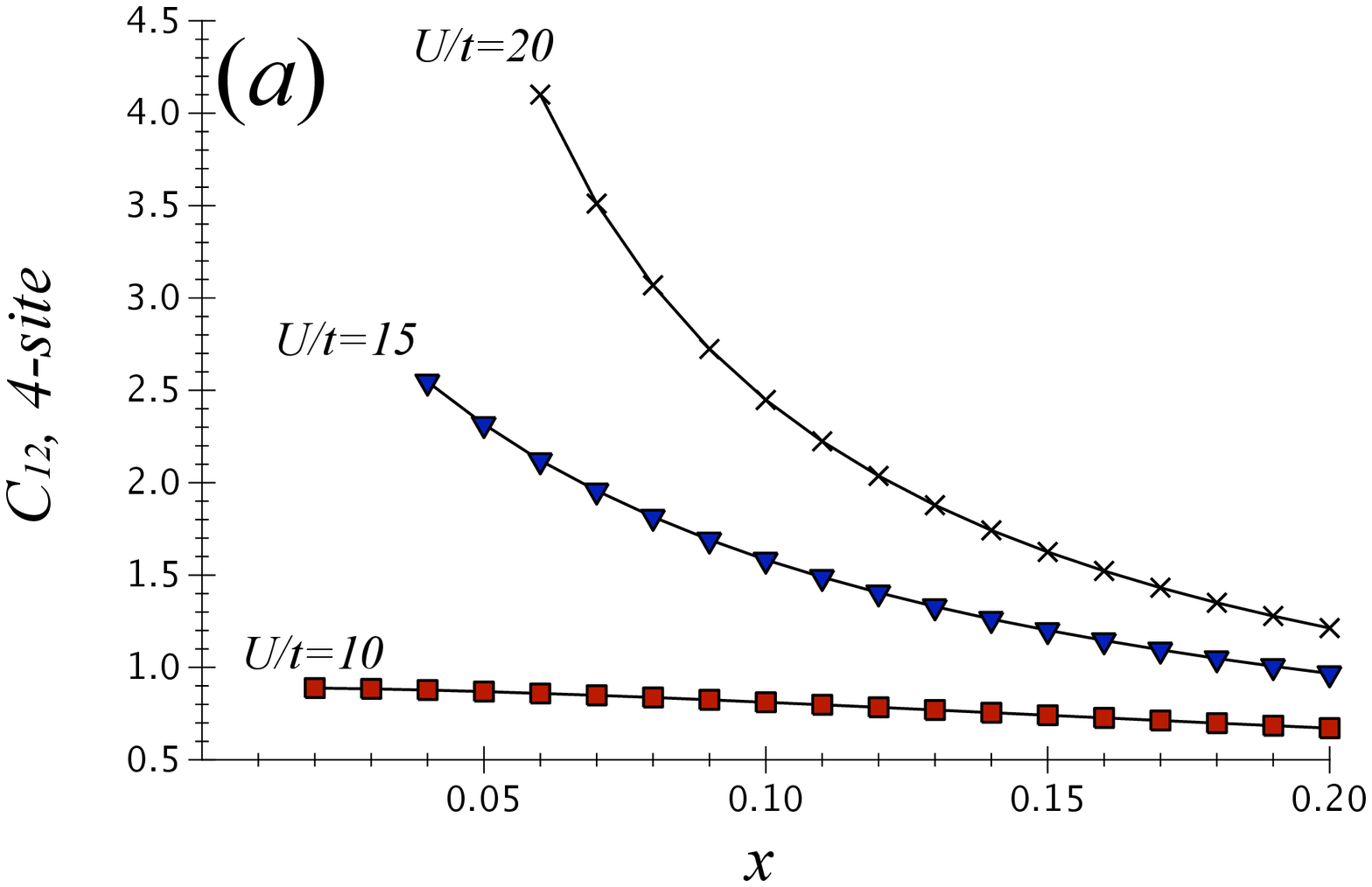}
\includegraphics[width=3in]{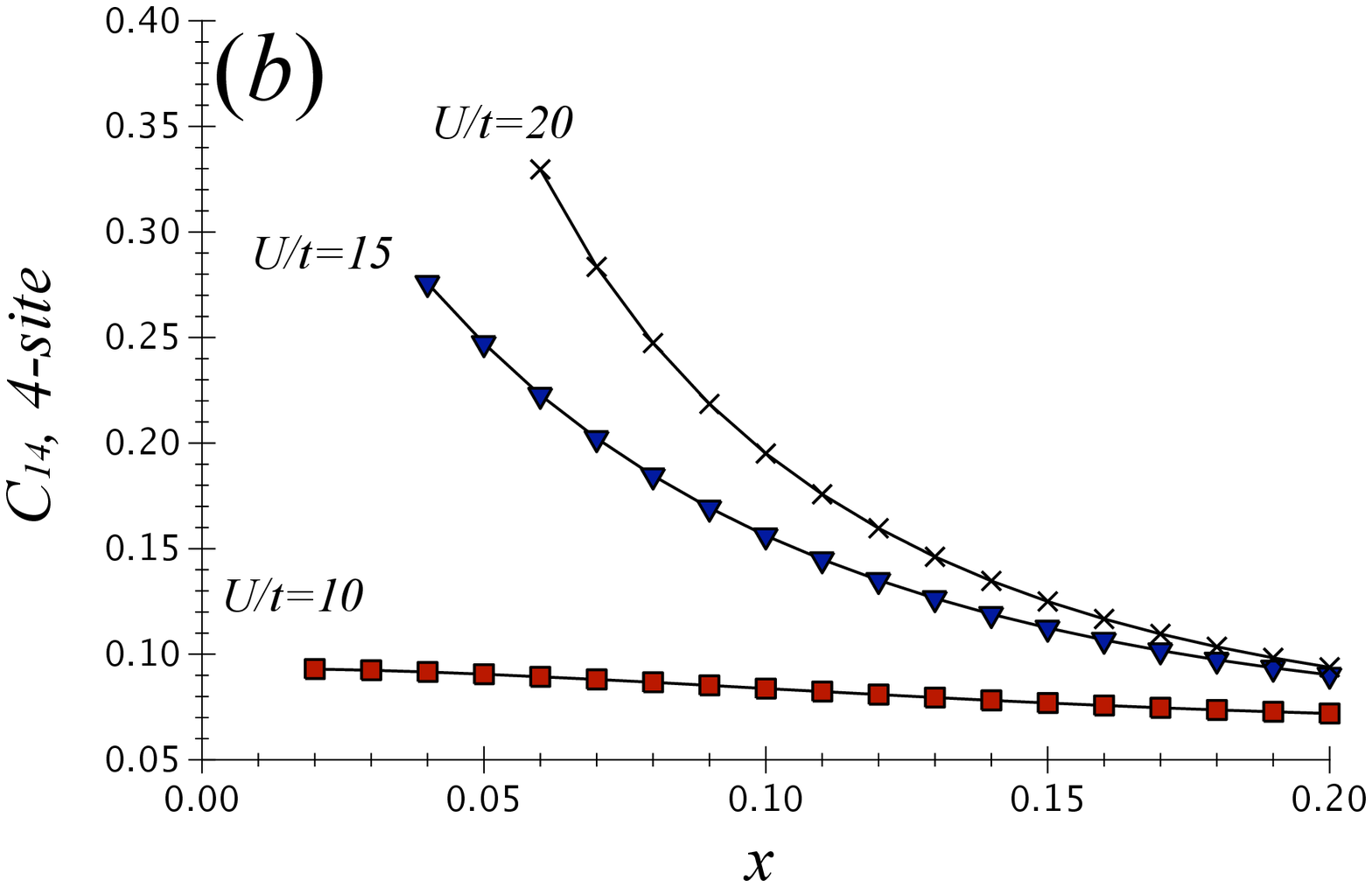}
\includegraphics[width=3in]{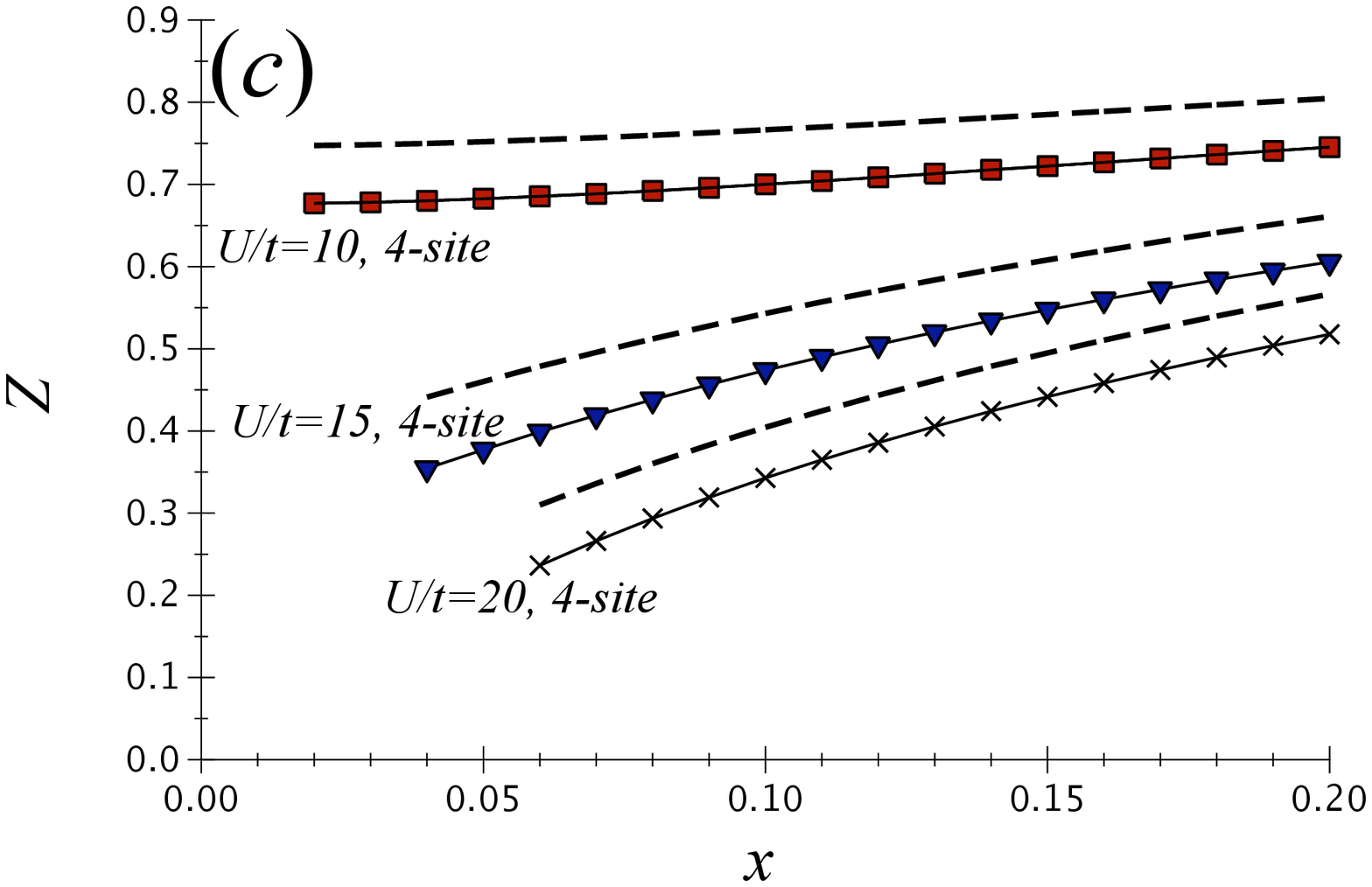}
\caption{\label{fig:para-4site} (Color online) (a) The holon-doublon correlation function for the nearest neighbor ($C_{12}$), (b) the next nearest neighbor ($C_{14}$), and
(c) the quasiparticle weight $Z$ as a function of doping $x$ at $U/t=10,15,20$ in the paramagnetic state. The dashed lines represent the corresponding Gutzwiller factors $g_t$ for each case.}
\end{figure}

In our cluster slave-spin method, we can obtain the paramagnetic state by finding the solution of $\Delta = 0$.
Figs. \ref{fig:para-2site} and \ref{fig:para-4site} plot the results obtained by the two- and four-site clusters as a function of doping $x$ at several different values of $U$.
We notice that the paramagnetic state can not exist as $U/t$ exceeds a doping-dependent critical value.
In the two-site cluster, no paramagnetic state exists for $x<0.02$ at $U/t=20$, and in the four-site cluster, no paramagnetic state exists for 
$x<0.02$ at $U/t=10$, $x<0.04$ at $U/t=15$, and $x<0.06$ at $U/t=20$ respectively.

It is interesting to also note that at large $U$ ($U/t=20$), $C_{12}$ in the paramagnetic state becomes greater than that in the AFM state, shown in Fig. \ref{fig:para-2site}(a) as 
$x<0.06$ for the two-site cluster and in Fig. \ref{fig:para-4site}(a) as $x<0.07$ for the four-site cluster.  
Moreover, in Figs. \ref{fig:para-2site}(b) and \ref{fig:para-4site}(c), 
we observe that the quasiparticle weight $Z$ is reduced for the two-site cluster, even more for the four-site, with respect to the single-site result of 
the Gutzwiller factor given in Eq. \ref{gtm0}. 
This reduction alleviates the shortcomings of the Gutzwiller approximation that gives a finite $Z$ at zero doping according to Eq. \ref{gtm0} even for very large $U/t$ 
where the state should be a Mott insulator.  
Therefore, even for a paramagnetic state, our cluster slave-spin method can have better results than the Gutzwiller approximation, and we can systematically improve the result 
by enlarging the cluster size. 

We can now conclude that the holon-doublon correlation is the essential part of the inter-site correlation in the strong-coupling regime where the Mott physics dominates, 
regardless of the antiferromagnetism. 
It is noted that although the four-site cluster captures the trend of the holon-doublon binding getting weaker at large doping, it still can not address critical issues regarding the 
physics of holon-doublon binding, e.g., whether there exists a quantum critical point associated with the unbinding of holon-doublon pairs.

\section{Summary and Discussion}
\begin{figure}
\includegraphics[width=3.5in]{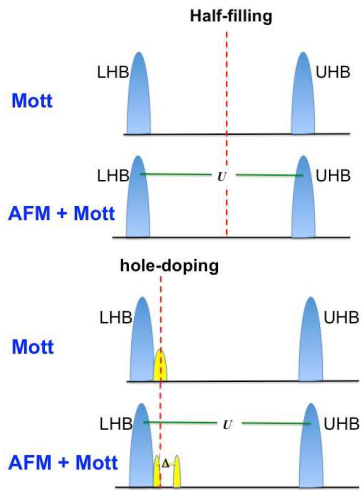}
\caption{\label{fig:phase} (Color online) Scenario for the AFM state in the Hubbard model. The red dashed line refers to the position of the chemical potential. 
The yellow areas near the chemical potential represent the coherent spectral weight of the electrons, and the blue areas marked as UHB (upper Hubbard band) and LHB (lower Hubbard band) 
represent the incoherent spectral weight of the electrons.}
\end{figure}
In summary, we have derived the formalism to study the density wave states using the cluster slave-spin method, and we have employed this method to investigate the antiferromagnetic (AFM)
state in the single band Hubbard model as functions of the on-site interaction $U$ and the doping $x$.
It is important to recognize that there are two distinct gaps predicted in this paper by the cluster slave-spin method.
The first one is the Hubbard (charge) gap which has been shown to scale with $U$ in the slave-spin method.\cite{yu2011,komijani2017} 
The second one is the the AFM gap $\Delta$, and our results have shown that $\Delta$ scales with $U$ in the weak coupling limit 
and goes like $t^2/U$ in the strong coupling regime. 
These two gaps come from fundamentally different mechanisms. 
The Hubbard gap is purely from the energy cost due to the doubly-occupied sites, thus it scales with $U$ and always exists regardless whether the system is in a symmetry-broken phase or not.
On the other hand, the AFM gap $\Delta$ is the energy gap induced by the AFM order in the coherent quasiparticles whose physical properties usually scale with $O(t/U)$
in the strong coupling limit.
As a result, $\Delta$ naturally scales with $t^2/U$ in the strong coupling regime as observed in our calculations.

Our scenario for the AFM state in the Hubbard model is schematically summarized in Fig. \ref{fig:phase}. 
For a Mott insulator at half-filling, there is no coherent quasiparticles near the Fermi energy ($Z=0$). 
Consequently, there is no AFM gap $\Delta$ even as the system is in the AFM state, and the Hubbard gap is the only gap in the Mott insulator.
For a doped Mott insulator, in addition to the Hubbard gap that remains roughly the same, 
the coherent quasiparticles can undergo a AFM transition and spin density wave (SDW) bands emerge.
The cluster slave-spin method predicts that
the AFM gap $\Delta$ separating the SDW bands scales with $U$ in the weak coupling limit and goes like $t^2/U$ in the strong coupling regime.
Our scenario is consistent with the recent variational Monte Carlo results in Ref. [\onlinecite{wuhk2017}] that 
there is no AFM gap at half filling and the AFM state.
The mechanism for the AFM state changes from the traditional mean-field picture to the superexchange mechanism as $U$ increases, 
and the cluster slave-spin method can capture such a crossover successfully within a single framework.
 
We also find that the physics related to the holon-doublon binding is an essential part in the inter-site fluctuations in the Hubbard model with or without broken symmetry.
In the four-site cluster, we are able to compute the holon-doublon correlation function for the nearest 
and next-nearest neighbors using the slave-spin representations for the holon and the doublon.
Our results suggest that the holon and the doublon tends to be bound with each other in the strong coupling regime, and the holon-doublon binding disappears either at small $U$ or at large 
doping $x$.
We expect that the results can be improved if more sites are introduced in the cluster.

\section{Acknowledgement}
We would like to thank E. Fradkin, A. MacDonald, H. Mathur, S. Mukherjee, and P. Phillips for fruitful discussions. 
W.C.L. is supported by a start up fund from Binghamton University. T.K.L. was partially supported by Taiwan Ministry of Science and Technology with Grant No.
105-2112-M-001-008, and the calculations were partially supported by the National Center for High Performance Computing in Taiwan.

\section{Appendix: Quasiparticle weight in the single-site $U(1)$ slave-spin method}
In the single-site $U(1)$ slave-spin method, the slave-spin Hamiltonian can be expressed as a four by four matrix whose basis can be written in terms of the Fock spaces of the 
Schwinger bosons: 
\beq
\vert 1\rangle &\equiv& a^\dagger_\uparrow a^\dagger_\downarrow\vert 0\rangle,\nn\\
\vert 2\rangle &\equiv& a^\dagger_\uparrow b^\dagger_\downarrow\vert 0\rangle,\nn\\
\vert 3\rangle &\equiv& b^\dagger_\uparrow a^\dagger_\downarrow\vert 0\rangle,\nn\\
\vert 4\rangle &\equiv& b^\dagger_\uparrow b^\dagger_\downarrow\vert 0\rangle.\nn\\
\label{es}
\eeq
The ground state wavefunction for the evaluation of the quasipartice weight $Z$ can therefore be expressed as
\be
\vert G\rangle = \alpha_1\vert 1\rangle + \alpha_2\vert 2\rangle + \alpha_3\vert 3\rangle + \alpha_4\vert 4\rangle,
\ee
subject to the normalization of $\sum_{n=1}^4 \vert \alpha_n\vert^2$. 
Because all the matrix elements in the slave-spin Hamiltonian are real numbers, we can assume $\alpha_{1,2,3,4}$ to be real numbers without loss of generality.
In the paramagnetic state, the system has the spin rotational symmetry. Using Eq. \ref{dresssb}, we have two equalities of
\beq
\langle G\vert S^z_\uparrow\vert G\rangle &=& \langle G\vert S^z_\downarrow\vert G\rangle,\nn\\
\langle G\vert \tilde{z}_\uparrow\vert G\rangle &=& \langle G\vert \tilde{z}_\downarrow\vert G\rangle,
\eeq
which leads to 
\be
\alpha_1\alpha_3 + \alpha_2\alpha_4 = \alpha_1\alpha_2 + \alpha_3\alpha_4.
\label{alphaall}
\ee
There are two different cases satisfying Eq. \ref{alphaall}.

{\it Case 1}: If $Z\neq 0$, the average spinon kinetic energy given in Eq. \ref{spinonke} is non-zero and so are $\alpha_{1,2,3,4}$. 
The solution to Eq. \ref{alphaall} in this case yields
\be
\alpha_2 = \alpha_3.
\label{alpha23}
\ee
Moreover, the holon and the doublon numbers are
\beq
h&=&\langle G\vert n^b_\uparrow n^b_\downarrow\vert G\rangle =\alpha_4^2,\nn\\
d&=&\langle G\vert n^a_\uparrow n^a_\downarrow\vert G\rangle =\alpha_1^2.
\eeq
Using the above equations and the relation between the holon and the doublon numbers given in Eq. \ref{hd}, we have
\beq
\alpha_1 &=& \sqrt{d},\nn\\
\alpha_4 &=& \sqrt{x + d}.
\eeq
Finally, using the constraint given in Eq. \ref{constraint}, we obtain
\beq
&&\langle G\vert S^z_\uparrow\vert G\rangle = \langle f^\dagger_{i\uparrow} f_{i\uparrow}\rangle-\frac{1}{2},\nn\\
&\to&\frac{1}{2}\big(\alpha_1^2 + \alpha_2^2 - \alpha_3^2 - \alpha_4^2\big)= \frac{1-x}{2}-\frac{1}{2} ,\nn\\
&\to&\alpha_1^2 - \alpha_4^2 = -x.
\label{alpha14}
\eeq
With Eq. \ref{alpha23}, the normalization condition becomes
\be
\alpha_1^2 + 2 \alpha_2^2 + \alpha_4^2 = 1.
\ee
Using the above equation and Eq. \ref{alpha14}, we obtain
\be
\alpha_2 = \sqrt{\frac{1-x}{2} - d}.
\ee

Now we have successfully expressed $\alpha_{1,2,3,4}$ in terms of the doping $x$ and the doublon number $d$. It is then straightforward to show
\beq
Z&=&\vert\langle G\vert \tilde{z}_\uparrow\vert G\rangle\vert^2\nn\\
&=& \frac{1}{(\frac{1}{2})^2 -\langle G\vert S^z_\uparrow\vert G\rangle^2}\alpha_2^2(\alpha_1 + \alpha_4)^2\nn\\
&=& \frac{1-x-2 d}{1-x-\frac{(1-x)^2}{2}} \big(\sqrt{x+d} + \sqrt{d}\big)^2,
\eeq
which reproduces exactly the Gutzwiller factor in the paramagnetic state.

{\it Case 2}: At half-filling ($x=0$), the Mott insulating phase characterized by $Z=0$ occurs as $U>U_c$, where $U_c$ is the critical interaction strength for the Mott transition.
Because of $Z=0$, all the off-diagonal terms in the single-site slave-spin Hamiltonian are zero in the Mott insulating phase.
In this case, the four states in Eq. \ref{es} are the eigen states and the corresponding eigen energies are: $E_1 = U+\lambda$, $E_2=0$, $E_3=0$, $E_4 = - \lambda)$ respectively.
$\lambda$ is the Langrangian multiplier ro enforce the constraint given in. Eq. \ref{constraint}, and at half-filling, $\lambda \to 0^-$.
As a result, $\vert G\rangle$ is a linear combination of $\vert 2\rangle$ and $\vert 3\rangle$ and consequently $\alpha_1$ and $\alpha_4$ are both zero. 
This solution of $\vert G\rangle$ automatically obtains $Z=0$ as the self-consistent solution.
 
In conclusion, the single-site slave-spin method obtains the Gutzwiller factor $g_t$ exactly for any case with finite $Z$, 
and it can still capture the Mott insulating phase at the half-filling as $U > U_c$, which goes beyond the Gutzwiller approximation.


\end{document}